\documentclass[a4paper,11pt]{article}
\usepackage{pos}
\usepackage{wrapfig}
\usepackage{subfigure}
\usepackage{float}
\usepackage[compact]{titlesec} 
\titlespacing{\section}{0pt}{*1.5}{*1}
\usepackage{lineno}
\title{Baikal-GVD: status and first results}
\author[a]{A.D.~Avrorin}
\author[a]{A.V.~Avrorin}
\author[a]{V.M.~Aynutdinov}
\author[b]{Z.~Bard\'{a}\v{c}ov\'{a}}
\author[c]{R.~Bannasch}
\author[d]{I.A.~Belolaptikov}
\author[d]{V.B.~Brudanin}
\author[e]{N.M.~Budnev}
\author[a]{G.V.~Domogatsky}
\author[a]{A.A.~Doroshenko}
\author[e]{A.N.~Dyachok}
\author[a]{Zh.-A.M.~Dzhilkibaev}
\author[d]{V.~Dik}
\author[b,d]{R.~Dvornick\'{y}}
\author[b]{E.~Eckerov\'{a}}
\author[d]{T.V.~Elzhov}
\author[f]{L.~Fajt}
\author[g]{S.V.~Fialkovski}
\author[e]{A.R.~Gafarov}
\author[a]{K.V.~Golubkov}
\author[d]{N.S.~Gorshkov}
\author[e]{T.I.~Gress}
\author[d]{R.A.~Ivanov}
\author[d]{M.S.~Katulin}
\author[c]{K.G.~Kebkal}
\author[c]{O.G.~Kebkal}
\author[d]{E.V.~Khramov}
\author[d]{M.M.~Kolbin}
\author[d]{K.V.~Konischev}
\author[h]{K.A.~Kopa\'{n}ski}
\author[d]{A.V.~Korobchenko}
\author[a]{A.P.~Koshechkin}
\author[i]{V.A.~Kozhin}
\author[a]{M.K.~Kryukov}
\author[d]{M.V.~Kruglov}
\author[g]{V.F.~Kulepov}
\author[a]{M.B.~Milenin}
\author[e]{R.R.~Mirgazov}
\author[d]{D.V.~Naumov}
\author[d]{V.~Nazari}
\author[h]{W.~Noga}
\author[a]{D.P.~Petukhov}
\author[d]{E.N.~Pliskovsky}
\author[j]{M.I.~Rozanov}
\author[d]{V.D.~Rushay}
\author[e]{E.V.~Ryabov}
\author*[a]{G.B.~Safronov}
\author[d]{B.A.~Shaybonov}
\author[a]{M.D.~Shelepov}
\author[b,d,f]{F.~\v{S}imkovic}
\author[i]{A.V.~Skurikhin}
\author[d]{A.G.~Solovjev}
\author[d]{M.N.~Sorokovikov}
\author[f]{I.~\v{S}tekl}
\author[d]{E.O.~Sushenok}
\author[a]{O.V.~Suvorova}
\author[e]{V.A.~Tabolenko}
\author[e]{B.A.~Tarashansky}
\author[d]{Y.V.~Yablokova}
\author[c]{S.~Yakovlev}
\author[a]{D.N.~Zaborov}

\affiliation[a]{Institute for Nuclear Research, Russian Academy of Sciences, Moscow, Russia}
\affiliation[b]{Comenius University, Bratislava, Slovakia}
\affiliation[c]{EvoLogics Gmbh, Berlin, Germany}
\affiliation[d]{Joint Institute for Nuclear Research, Dubna, Russia}
\affiliation[e]{Irkutsk State University, Irkutsk, Russia}
\affiliation[f]{Czech Technical University in Prague, Prague, Czech Republic}
\affiliation[g]{Nizhny Novgorod State Technical University, Nizhny Novgorod, Russia}
\affiliation[h]{Institute of Nuclear Physics of Polish Academy of Sciences (IFJ~PAN), Krak\'{o}w, Poland}
\affiliation[i]{Moscow State University, Moscow, Russia}
\affiliation[j]{St.~Petersburg State Marine Technical University, St.Petersburg, Russia}

\emailAdd{grigorybs@gmail.com}
\abstract{Baikal-GVD is a cubic-kilometer scale deep-underwater neutrino detector being constructed in Lake Baikal. It is designed to detect neutrinos from $\sim$100 GeV to  multi-PeV energies and beyond. Detector deployment began in Spring 2015. Since April 2020 the detector includes seven 8-string clusters carrying in total 2016 optical modules located at depths from 750 to 1275 meters. By the end of the first phase of detector construction in 2024 it is planned to deploy 15 clusters, reaching the effective volume for high-energy cascade detection of 0.75 km$^3$. The design and status of the Baikal-GVD detector and first results of data analysis are presented in this report.}

\FullConference{%
  40th International Conference on High Energy physics - ICHEP2020\\
  July 28 - August 6, 2020\\
  Prague, Czech Republic (virtual meeting)
}

\begin{document}

\maketitle

\section{Introduction}
Neutrinos produced in high-energy extraterrestrial acceleration sites travel large distances not scattered by either interstellar or intergalactic medium and undeflected by cosmic magnetic fields. Information about neutrino direction, energy and time of arrival at Earth combined with results of gamma-ray, optical, cosmic ray and gravitational wave observations can constrain models of high-energy cosmic particle production mechanisms. Back in 1960 M.A. Markov proposed to instrument natural water reservoirs with sparse arrays of photodetectors to detect upgoing charged particles by means of Cerenkov radiation \cite{markov}. Neutrino interactions near or inside the instrumented water volume can be detected by reconstructing the cascade of charged particles accompanying the interaction. A larger detection range is accessible for $\nu_{\mu}$ and partly $\nu_{\tau}$ interactions mediated by W$^{\pm}$ which are accompanied by the muon track due to long range of high energy muons. This technique allows to increase the sensitive volume of the detector to the order of 1~km$^3$ making it a good choice for the search of small flux of high-energy astrophysical neutrinos. There are three large-scale neutrino telescopes operating presently: ANTARES telescope in the Mediterranean sea, IceCube in the Antarctic Ice Sheet, and Baikal-GVD in Lake Baikal. In addition work on constructing ANTARES successor KM3NET telescope is ongoing in the Mediterranean. IceCube was the first detector to reach instrumented volume of 1km$^3$. Sensitivity of these devices extends from 10's - 100's of GeV to multi-PeV neutrino energies and beyond with lower bound depending on density of photodetectors and ice or water light absorption length. In this energy range most of neutrinos reaching the detectors are produced in cosmic ray interactions with Earth's atmosphere. The IceCube experiment demonstrated an excess of neutrino events over the atmospheric neutrino spectrum starting from $\sim$100 TeV in 2011 - 2013 \cite{ic_diff_discov}, in 6 year data analysis released in 2016 the significance of an excess has reached 8$\sigma$ \cite{ic_review}. Presently these events are attributed to neutrinos produced in high-energy processes in remote cosmic objects, however any indications of astrophysical neutrino sources have not yet reached the 5$\sigma$ discovery level. An important step towards understanding of cosmic neutrino origin was the detection of neutrino alert event associated with blazar TXS0506+056 \cite{ic_txs}, analysis of IceCube events prior to alert had shown 3.5$\sigma$ evidence of neutrino emission from that blazar direction \cite{ic_txs_prior}. The recent point source search with 10 year data of IceCube track-like events indicate 3.3$\sigma$ inconsistency with background for the full Nothern Hemisphere source catalog \cite{ic_10y}. Moreover recent studies indicate 4.1$\sigma$ association of IceCube track-like events with radio-bright blazars selected on the basis of very long baseline interferometry observations \cite{plavin2}. 

Work on construction of the neutrino telescope in Lake Baikal began in 1980. In the mid-1990s first underwater neutrinos were detected consecutively at the NT-36 and NT-96 neutrino telescopes with 36 and 96 photodetectors respectively \cite{nt96_perf,nt96_neutrino}. Work on construction of the Baikal-GVD telescope of the cubic-kilometer scale began in 2011, and the first demo cluster named Dubna was commissioned in 2015. Since April 2020, the telescope includes seven clusters carrying 2016 optical modules (OMs). The present work reports on the design and status of the Baikal-GVD detector and on first results of data analysis.

\section{Baikal-GVD detector}
\begin{wrapfigure}[20]{r}{0.4\textwidth}
\centering
    \includegraphics[width=0.4\textwidth]{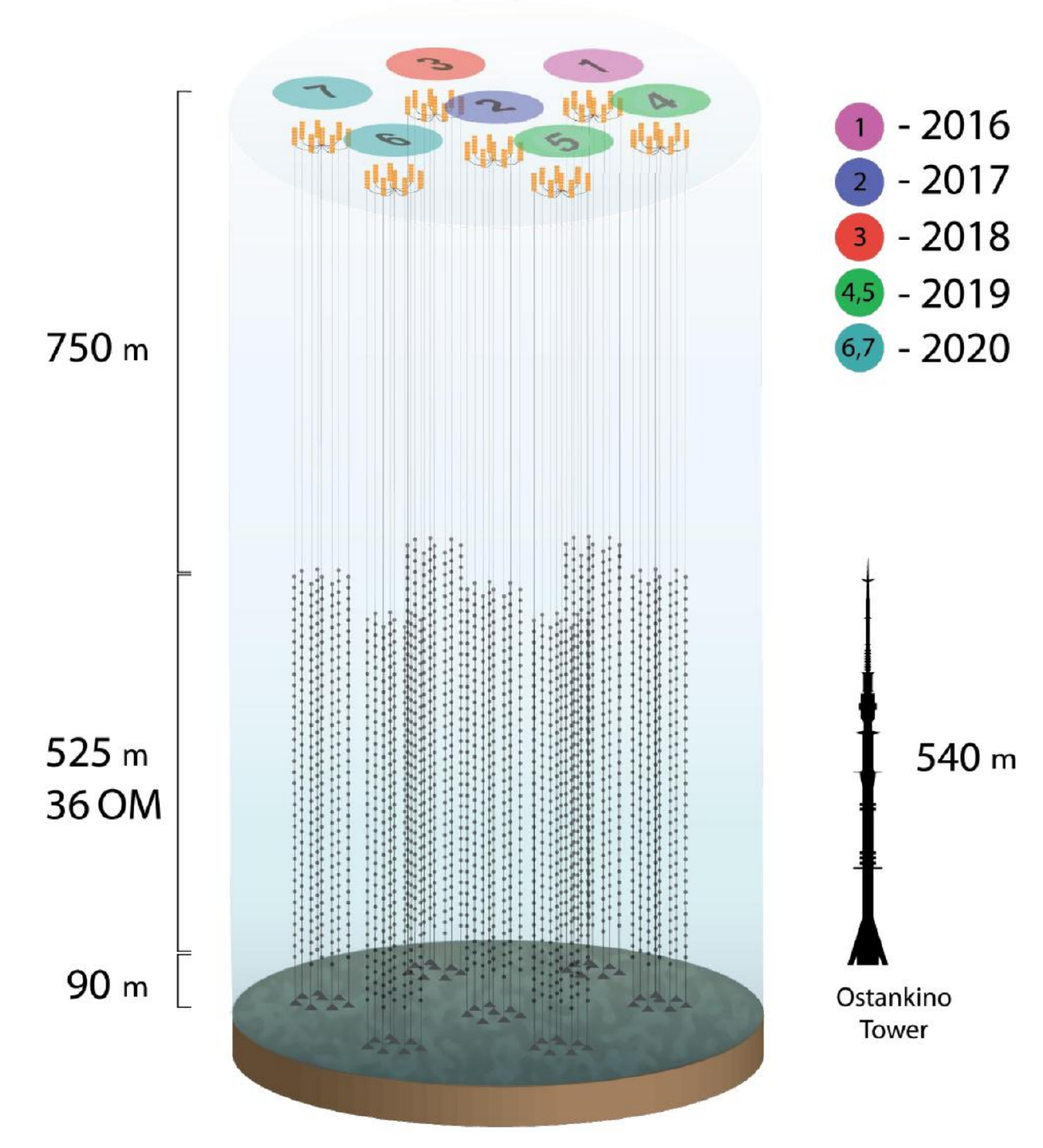}
    \caption{General view of Baikal-GVD detector in 2020. Cluster radius is 60 m and the distance between neighbouring cluster centers is $\sim$ 300 m. A yearly progression of the detector deployment is shown in the legend.}
    \label{fig:det}
\end{wrapfigure}

The experiment site is located in the southern part of Lake Baikal 3.6 km away from shore where the lake depth is nearly constant at 1366-1367 m. Telescope consists of independent structural units - clusters (Fig.~\ref{fig:det}). Cluster consists of 8 strings each carrying 36 optical modules (OM) and calibration systems. 
OMs are located at depths from 750 to 1275 (m) with a vertical step of 15 m. Optical module is a glass sphere which hosts photomultiplier tube (PMT) Hamamatsu R7081-100 with a 10-inch hemispherical photocathode oriented towards the lake floor along with various sensors, readout and HV control electronics, and calibration LED. OM readout is organised in sections each including 12 OMs. Readout controllers are housed in a section's Central Module (CM) in which digitisation of PMT signal occurs with a step of 5 ns. Time of signal arrival and charge deposited in PMT is derived from the pulse shape analysis. CM generates local trigger signal if pulses with deposited charge above adjustable thresholds $Q_{high}$ and $Q_{low}$ coincident within 100 ns time window are found in adjacent modules within one section. 
Typical values for $Q_{high}$ are 3-5 photoelectrons (p.e.) and 1-2 p.e. for $Q_{low}$ depending on the cluster and season. If a local trigger is generated within one of CM an event frame of 5~$\mu$s is read out from all CMs of the cluster and sent to the shore center via optoelectric cable attached to a central cluster control module. Data are further transmitted to JINR (Dubna) for event reconstruction and long-term storage.
Baikal water currents cause devations of topmost OMs up to 50 m from their median position with average speed of 0.5 cm/s. For precise OM positioning each string carries system of acoustic modems, a cross-poll of modems if performed each 1 - 6 minutes. OM coordinates are reconstructed with the precision better than 20~cm \cite{acoustic}. Single cluster time calibration is performed with the precision of 2.5 ns using OM integrated LEDs and LED matrices which produce light flashes propagating to up to 100 m radius~\cite{tcalib}. A system of three technological strings located between clusters carries 5 dedicated lasers producing isotropical flashes which are used for intercluster calibration and water properties monitoring. 
The light absorption length of Baikal water amounts to $\sim$22~m.
The PMT noise rate is dominated by chemiluminescence, with substantial seasonal variations. Noise pulses at the level of 1 p.e. are produced at the rate of 20-50 kHz in quiet period in April - June and up to > 100 kHz at topmost OMs for some periods in the rest of the year \cite{baikal_noise_icrc}. The noise rate is monitored and is taken into account in the detector Monte-Carlo (MC) simulations.

\section{Physics results}

\begin{figure}[ht]
\centering
 \subfigure[]
 {
    \includegraphics[width=0.4\textwidth]{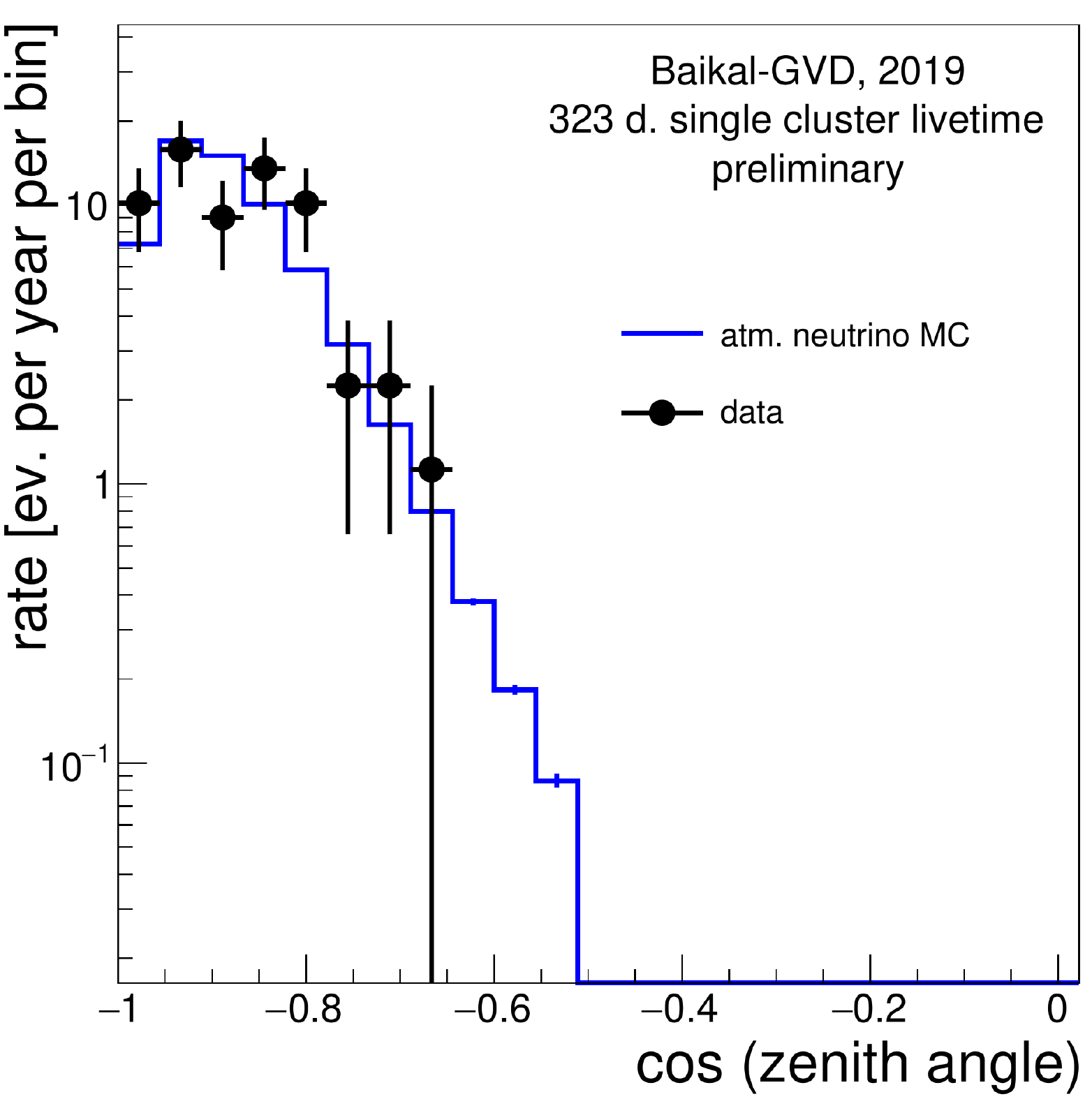}
    }
 \subfigure[]
 {
    \includegraphics[width=0.17\textwidth]{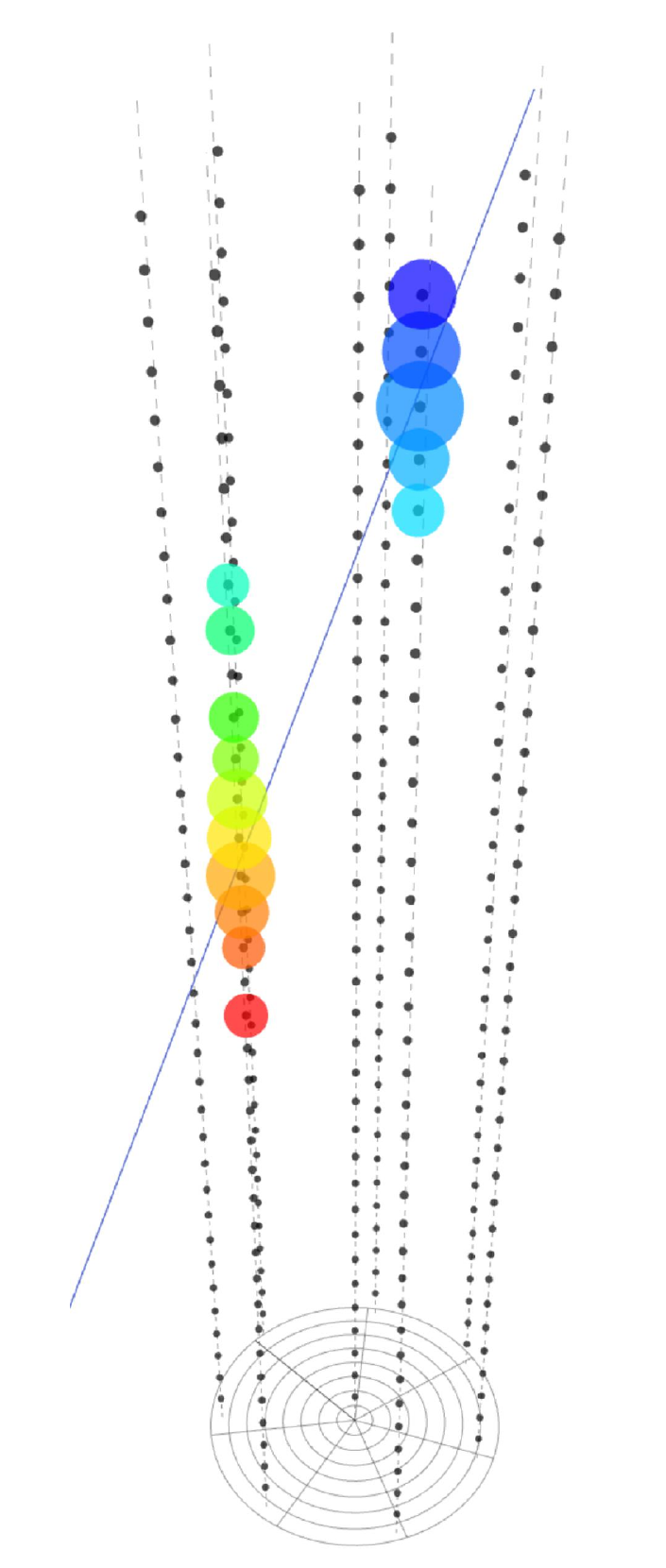}
	}
 \subfigure[]
 {	
    \includegraphics[width=0.17\textwidth]{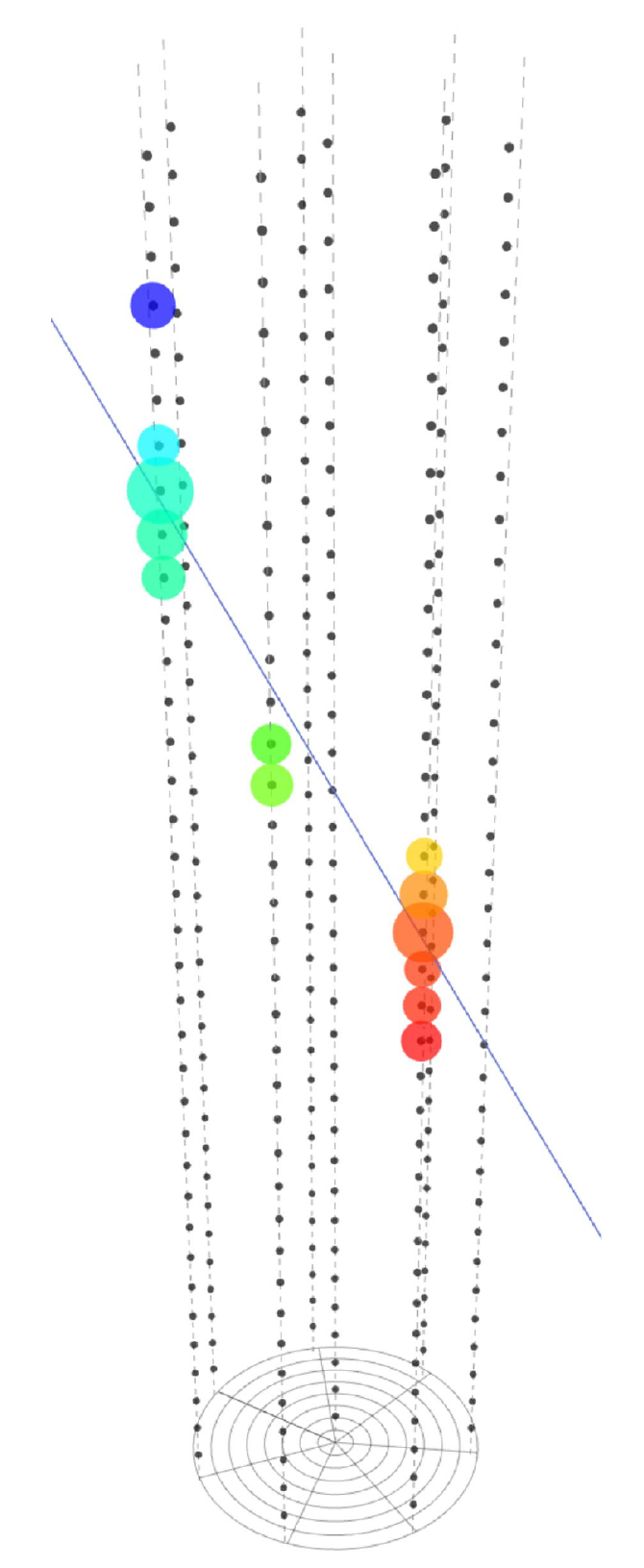}
	}
    \caption{(a): Zenith angle distribution of 57 neurino candidate events selected in April - June 2019. An expectation for backgound atmospheric muons was estimated to < 1 events and thus is not shown. (b): Neutrino candidate reconstructed on June 11th 2019 in cluster 1, $\Theta_{zenith}=161.8^{\circ}$. (c): Neutrino candidate reconstructed on June 18th 2019 in cluster 5, $\Theta_{zenith}=148.0^{\circ}$. On event displays early hits are shown as red and late hits in blue. }
    \label{fig:neutrino}
\end{figure}

Presently the data are analysed for each of 7 clusters idependently what is referred to as single cluster analysis while the multi cluster analysis is in development.
The muon track reconstruction is performed in two stages. At first stage PMT pulses are clustered using the time and distance constraints with respect to the seed pulse with high charge deposition. A preliminary muon track parameters estimation is performed and outlier pulses are removed. As a results of this procedure the collection of pulses with average noise contribution at the level of 1\% is selected.
At the second stage the muon track parameters are reconstructed by means of minimisation of the quality function $Q=\chi^{2}(t) + w*f(q,r)$ where $\chi^{2}(t)$ is the chi-square sum of time residuals with respect to direct Cerenkov light from the muon, $f(q,r)$ is the sum of products of charges deposited in OMs and their distances from the track and $w$ is the relative weight of the second term. 
\begin{wrapfigure}[18]{l}{0.4\textwidth}
\centering
  \includegraphics[width=0.4\textwidth]{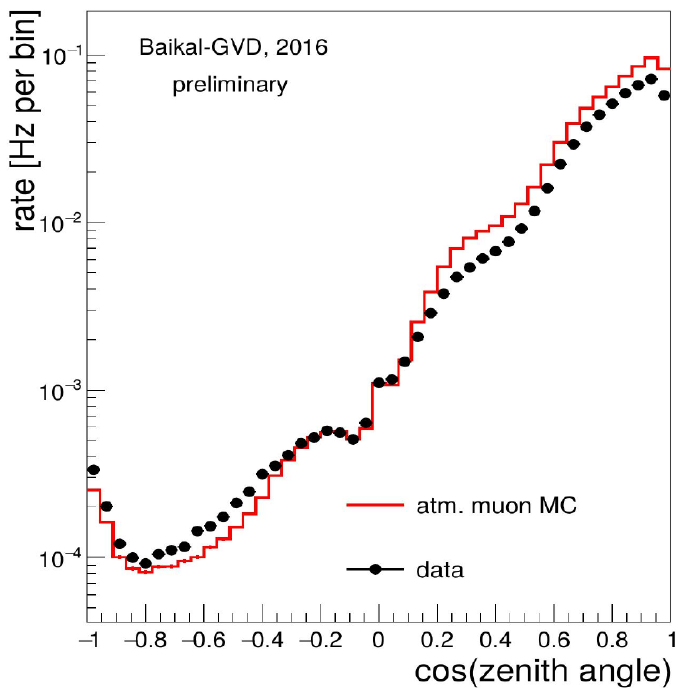}
  \caption{Reconstructed muon zenith angle distribution in cluster 1 of 2016 season. Data are compared to MC simulation based on CORSIKA event generator~\cite{corsika}.}
\label{fig:muons}
\end{wrapfigure}
Events with at least 6 pulses at 2 strings are used for the reconstruction. 
Distribution of zenith angle for tracks reconstructed in quiet period of season 2016 with minimal quality requirements is shown at Fig.~\ref{fig:muons}. A few~\% of events are reconstructed as upgoing muons forming a background to upgoing neutrino events exceeding the atmospheric neutrino flux by a factor $10^5-10^6$. To suppress the background from misreconstructed tracks a set of cuts on 13 reconstructed track parameters is used with the most powerful variable being $Q/ndf$. 
%Most powerful variable found to be $Q/ndf$. 
For the single cluster analysis the zenith angle is constrained to $\Theta_{zenith} > 120^{\circ}$ to ensure sufficient length of the track for high quality reconstruction. Median angular resolution for events passing the analysis cuts is $\sim$1.0$^{\circ}$. 
The developed cuts were applied to a collection of runs from April 1st to June 30th 2019 (quiet period). 
The total live time of selected sample is 323 days of single cluster data-taking. A total of 57 neutrino candidate events were selected while the expectation from atmospheric neutrino MC simulation is 54.3 events (Fig.~\ref{fig:neutrino}). 
\newpage

Cascade event energy and direction are reconstructed using maximum likelihood technique where time and charge of pulses are taken into account \cite{baikal_casc_icrc}. For the cascade of energy 100 TeV the uncertainty of cascade direction reconstruction amounts to $\sim$4$^{\circ}$ while the energy reconstruction uncertainty is $\delta E/E \sim 30\%$. 
The telescope effective volume for detecting high-energy cascades was 0.25 km$^{3}$ in 2019 and became 0.35 km$^3$ in 2020. Based on the IceCube fit of the astrophysical neutrino spectrum \cite{ic_spectrum_modern} 0.6 neutrino events are expected per year per cluster or 4.2 events per year for detector configuration in 2020. The high-energy cascade search was performed using the data of years 2015, 2016, 2018 and 2019. The sum live time for the used data is 2294 days of single cluster data-taking. 
Events with reconstructed energies above 100 TeV with at least 19 fired OMs were selected as neutrino candidate events. In 2015-2019 data twelve such events were found, all of these events were reconstructed as downgoing. In addition one upgoing event with 19 hits and reconstructed energy $\sim$91 TeV was found in 2019 data. 

\begin{wrapfigure}[18]{r}{0.45\textwidth}
\centering
  \includegraphics[width=0.45\textwidth]{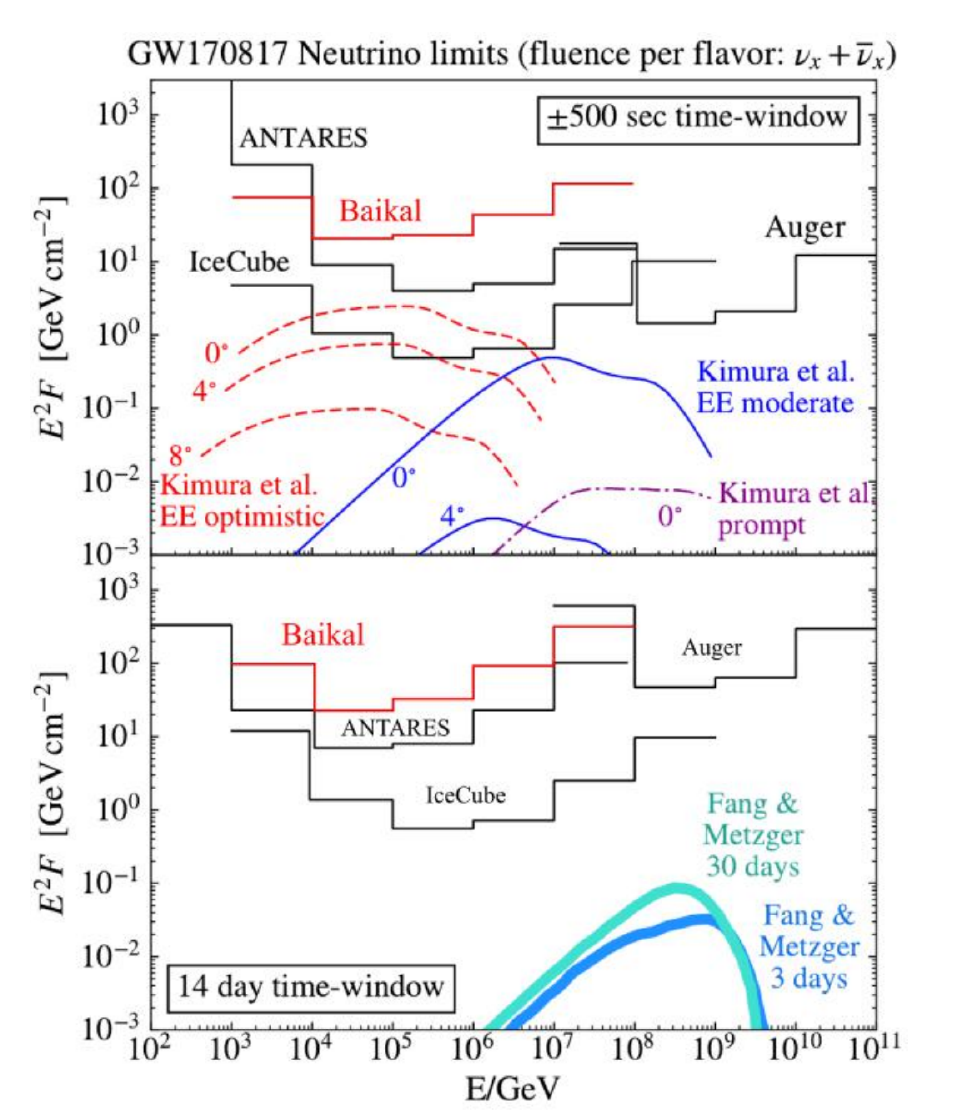}
  \caption{Limits on the neutrino flux from GW170817 event. Figure is taken from~\cite{gw_gvd}.}
\label{fig:gw}
\end{wrapfigure}

On August 17, 2017 the LIGO and VIRGO detectors of gravitational waves recorded a signal named GW170817 which was folllowed by gamma and optical bursts \cite{gw170817}. 
This event was associated with the merger of neutron stars in the NGC 4993 galaxy. 
The ANTARES and IceCube telescopes, the Pierre-Auger observatory and the Super-Kamiokande detector performed searches for neutrino associated with this event \cite{gw170817_neutrino,gw_superk}. 
In Baikal-GVD the search for neutrinos from GW170817 was performed in the cascade channel \cite{gw_gvd}. In 2017 Baikal-GVD consisted of two clusters. At the time of the event the NGC 4993 galaxy was approximately at the horizon ($\Theta_{zenith}=93.3^{\circ}$). The search was conducted in the time window of $\pm$500 s and $\pm$14 days around the event to take into account all neutrino production models. The requirement on the number of fired optical modules was relaxed but no event in the direction of of NGC 4993 was detected. A limit  on the neutrino flux at a 90\% C.L. was set under the assumption of $E^{-2}$ spectrum and identical fluxes of all neutrino flavors (Fig.~\ref{fig:gw}).

\section{Conclusions}
The Baikal-GVD deployment is ongoing with a pace of 2 clusters per year. First results from uncompleted detector are available. A first set of track-like neutrino candidate events obtained using the data from April - June of 2019 was presented. An acceptable agreement of track-like neutrino event rate with MC expectation was demonstrated. A set of twelve downgoing high energy cascade events with estimated energy above 100 TeV and an upgoing cascade event with estimated energy of $\sim$91 TeV were selected in 2015-2019 data. A limit on neutrino flux from neutron star merger event GW170817 was set.  

This work is supported by the Ministry of Science and Higher Education of Russian Federation under the contract 075-15-2020-778 in the framework of the Large scientific projects program within the national project "Science".
We acknowledge the support from RFBR grant 20-02-00400 and grant 12-29-11-029, as well as support by the JINR young scientist and specialist grant 20-202-09. We also acknowledge the technical support of JINR staff for the computing facilities (JINR cloud).

%\begin{thebibliography}{99}
%\bibitem{...}
%....

%\end{thebibliography}

\end{document}